\documentstyle[12pt]{article}
\begin{document}

\vspace*{4.5cm}

\noindent
{\bf NEUTRINO OSCILLATIONS: SOME THEORETICAL IDEAS\footnote{Talk given
at Orbis Scientiae Conference, Coral Gables, FLA, Dec. 16-19, 1999}}

\vspace{1cm}

\hspace{0.75in} Stephen M. Barr 

\vspace{0.5cm}

\hspace{0.75in} Bartol Research Institute 

\hspace{0.75in} University of Delaware 

\hspace{0.75in} Newark, DE 19711 

\vspace{1cm}

\noindent
{\bf INTRODUCTION}

\vspace{0.5cm}

Over the years, and especially since the discovery of the large
mixing of $\nu_{\mu}$ seen in atmospheric neutrino
experiments, there have been numerous models of neutrino masses
proposed in the literature. In the last two years alone, as
many as one hundred different models have been published.
One of the goals of this talk is to give a helpful classification
of these models. Such a classification is possible because
in actuality there are only a few basic ideas that underlie the
vast majority of published neutrino mixing schemes. 
After some preliminaries, I give a classification of three-neutrino
models, and then in the last part of the talk I discuss in more detail
one category of models --- those with ``lopsided" charged-lepton mass 
matrices. Finally, I talk about a specific very predictive model based on
lopsided mass matrices that I have worked on with Albright and Babu.

\vspace{1.0cm}

\noindent
{\bf THE DATA}

\vspace{0.5cm}

There are four indications of neutrino mass that guide
recent attemps to build models: (1) the solar neutrino problem,
(2) the atmospheric neutrino anomaly, (3) the LSND experiment, and
(4) dark matter. Several excellent reviews of the evidence for
neutrino mass have appeared recently.$^1$

(1) The three most promising solutions to the solar 
neutrino problem are based on neutrino mass. These are the
small-angle MSW solution (SMA), the large-angle MSW solution (LMA), and 
the vacuum oscillation solution (VO). All these solutions involve
$\nu_e$ oscillating into some other type of neutrino ---
in the models we shall consider predominantly $\nu_{\mu}$. In the
SMA solution the mixing angle and mass-squared splitting between
$\nu_e$ and the neutrino into which it oscillates are roughly
$\sin^2 2 \theta \sim 5.5 \times 10^{-3}$ and $\delta m^2 
\sim 5.1 \times 10^{-6} eV^2$. 
For the
LMA solution one has $\sin^2 2 \theta \sim 0.79$, and $\delta m^2
\sim 3.6 \times 10^{-5} eV^2$. (The numbers are best-fit
values from a recent analysis.$^2$) And for the VO solution 
$\sin^2 2 \theta \sim 0.93$, and 
$\delta m^2 \sim 4.4 \times 10^{-10} eV^2$. (Again, these are best-fit
values from a recent analysis.$^3$)

(2) The atmospheric neutrino anomaly strongly implies that $\nu_{\mu}$
is oscillating with nearly maximal angle into either $\nu_{\tau}$
or a sterile neutrino, with the data preferring the former
possibility.$^4$ One has $\sin^2 2 \theta \sim 1.0$, and $\delta m^2
\sim 3 \times 10^{-3} eV^2$.

(3) The LSND result, which would indicate a mixing between $\nu_e$
and $\nu_{\mu}$ with $\delta m^2 \sim 0.1 - 1 eV^2$ is regarded with more 
skepticism for two reasons.
The experimental reason is that KARMEN has failed to corroborate
the discovery, though KARMEN has not excluded the entire LSND region.
The theoretical reason is that to account for
the LSND result and also for both the solar and atmospheric anomalies
by neutrino oscillations would require three quite different
mass-squared splittings, and that can only be achieved with
{\it four} species of neutrino. This significantly complicates
the problem of model-building. In particular, it is regarded as
not very natural, in general, to have a fourth sterile neutrino
that is extremely light compared to the weak scale. For these
reasons, the classification given in this talk will assume that
the LSND results do not need to be explained by neutrino
oscillations, and will include only three-neutrino models.

(4) The fourth possible indication of neutrino mass is the existence of
dark matter. If a significant amount of this dark matter is in
neutrino mass, it would imply a neutrino mass of order several eVs.
In order then to achieve the small $\delta m^2$'s needed to
explain the solar and atmospheric anomalies one would have to assume 
that $\nu_e$, $\nu_{\mu}$ and $\nu_{\tau}$ were nearly degenerate.
We shall not focus on such models in our classification, which
is primarily devoted to models with ``hierarchical" neutrino masses. 
However, in most models with nearly degenerate masses, the neutrino mass
matrix consists of a dominant piece proportional to the identity
matrix and a much smaller hierarchical piece. Since the piece proportional
to the identity matrix would not by itself give oscillations, such
models can be classified together with hierarchical mass models in
most instances.

In sum, the models we shall classify are those which assume
(a) three flavors of neutrino that oscillate ($\nu_e$, $\nu_{\mu}$,
and $\nu_{\tau}$), (b) a hierarchical pattern of neutrino masses,
(c) the atmospheric anomaly explained by $\nu_{\mu}$-$\nu_{\tau}$
oscillations with nearly maximal angle, and (d) the solar
anomalies explained by $\nu_e$ oscillating primarily with $\nu_{\mu}$
with either small angle (SMA) or large angle (LMA, VO).

\vspace{1.0cm}

\noindent
{\bf MAJOR DIVISIONS}

\vspace{0.5cm}
 
There are several major divisions among models. One is
between models in which the neutrino masses arise through
the see-saw mechanism,$^5$ and those in which the neutrino masses are
generated directly at low energy. In see-saw models, there are
both left- and right-handed neutrinos. Consequently, there are
five fermion mass matrices to explain: the four Dirac mass
matrices, $U$, $D$, $L$, and $N$ of the up quarks, down quarks, 
charged leptons, and neutrinos, respectively, and the Majorana
mass matrix $M_R$ of the right-handed neutrinos. The four Dirac
mass matrices are all roughly of the weak scale, while $M_R$
is many orders of magnitude larger than the weak scale. After
integrating out the superheavy right-handed neutrinos,
the mass matrix of the left-handed neutrinos is given by
$M_{\nu} = - N^T M_R^{-1} N$. Typically, 
in see-saw models, the four Dirac mass matrices are closely
related to each other, either by grand unification or by flavor 
symmetries.
That means that in see-saw models neutrino masses and mixings are 
just one aspect of the larger problem of quark and lepton masses, and
are likely to shed great light on that problem, and perhaps even be
the key to solving it. On the other hand, in most see-saw models
$M_R$ is either not related or is tenuously related to the
Dirac mass matrices of the quarks and leptons. The freedom in
$M_R$ is the major obstacle to making precise predictions of
neutrino masses and mixings in most see-saw schemes.

In non-see-saw schemes, there are no right-handed neutrinos.
Consequently, there are only four mass matrices to
consider, the Dirac mass matrices of the quarks and charged leptons,
$U$, $D$, and $L$, and the Majorana mass matrix of the light
left-handed neutrinos $M_{\nu}$. Typically in such schemes $M_{\nu}$
has nothing directly to do with the matrices $U$, $D$, and $L$, but
is generated at low-energy by completely different physics.

The three most popular possibilities in recent models for generating
$M_{\nu}$ at low energy in a non-see-saw way are (a) triplet Higgs,
(b) variants of the Zee model,$^6$ and (c) R-parity violating terms in
low-energy supersymmetry. (a) In triplet-Higgs models, $M_{\nu}$ arises
from a renormalizable term of the form $\lambda_{ij} \nu_i \nu_j H_T^0$,
where $H_T$ is a Higgs field in the $(1,3, +1)$ representation
of $SU(3) \times SU(2) \times U(1)$. (b) In the Zee model, the Standard
Model
is supplemented with a scalar, $h$, in the $(1,1, +1)$ representation
and having weak-scale mass. 
This field can couple to the lepton doublets $L_i$ as $L_i L_j h$
and to the Higgs doublets $\phi_a$ (if there is more than one)
as $\phi_a \phi_b h$. Clearly it is not possible to assign
a lepton number to $h$ in such a way as to conserve it in both 
these terms. The
resulting lepton-number violation allows one-loop diagrams that 
generate a Majorana mass for the left-handed neutrinos. (c) In
supersymmetry the presence of such R-parity-violating terms in
the superpotential as $L_i L_j E^c_k$ and $Q_i D^c_j L_k$,
causes lepton-number violation, and allows one-loop diagrams that
give neutrino masses.

It is clear that in all of these schemes the couplings that give
rise to neutrino masses have nothing to do with the physics that
gives mass to the other quarks and leptons. While this allows
more freedom to the neutrino masses, it would from one point of view
be very disappointing, as it would mean that the observation of
neutrino oscillations is almost irrelevant to the burning question
of the origin of quark and charged lepton masses.

Another major division among models has to do with the kinds of symmetries
that constrain the forms of mass matrices and that, in some models,
relate different mass matrices to each other. There are two
main approaches: (a) grand unification, and (b) flavor symmetry. Many
models use both. 

(a) The simplest grand unified group is $SU(5)$.
In minimal $SU(5)$ there is one relation among the Dirac mass matrices,
namely $D= L^T$, coming from the fact that the left-handed charged
leptons are unified with the right-handed down quarks in a 
$\overline{{\bf 5}}$, while the right-handed charged leptons and 
left-handed down quarks are unified in a ${\bf 10}$. In $SU(5)$
there do not have to be right-handed neutrinos, though they may
be introduced. In $SO(10)$, which in several ways is a very
attractive group for unification, the minimal model gives the relations
$N = U  \propto D = L$. In realistic models these relations
are modified in various ways, for example by the appearance
of Clebsch coefficients in certain entries of some of the mass matrices.
It is clear that unified symmetries are so powerful that very
predictive models are possible. Most of the published models which give
sharp predictions for masses and mixings are unified models.

(b) Flavor symmetries can be either abelian or non-abelian.
Non-abelian symmetries are useful for obtaining the 
equality of certain elements of the mass matrix, as in models
where the neutrino masses are nearly degenerate, and in the
so-called ``flavor democracy" schemes. Abelian symmetries
are useful for explaining hierarchical mass matrices through the
so-called Froggatt-Nielson mechanism.$^7$ The idea is that
different fermion multiplets can differ in charge under a $U(1)$ flavor 
symmetry that is spontaneously broken by some ``flavon" 
expectation value (or values), $\langle f_i \rangle$. Thus,
different elements of the fermion mass matrices would be suppressed by
different powers of $\langle f_i \rangle/M \equiv \epsilon_i \ll 1$,
where $M$ is the scale of flavor physics. This kind of
scheme can explain small mass ratios and mixings in the sense of
predicting them to arise at certain orders in the small quantities 
$\epsilon_i$. A drawback of such models compared to many
grand unified models, is that actual numerical predictions,
as opposed to order of magnitude estimates, are not possible.
On the other hand, models based on flavor symmetry involve less
of a theoretical superstructure built on top of the Standard Model
than do unified models, and could therefore be considered more
economical in a certain sense. Unified models put more in but get more
out than flavor symmetry.

\vspace{1.0cm}

\noindent
{\bf THE PUZZLE OF LARGE $\nu_{\mu}-\nu_{\tau}$ MIXING}

\vspace{0.5cm}

The most significant new fact about neutrino mixing is the largeness
of the mixing between $\nu_{\mu}$ and $\nu_{\tau}$
This comes as somewhat of a surprise from the point of view of both
grand unification and flavor symmetry approaches. Since grand
unification relates leptons to quarks, one might expect lepton 
mixing angles to be small like those of the quarks. In particular,
the mixing between the second and third family of quarks is given by
$V_{cb}$, which is known to be $0.04$. That is to be
compared to the nearly maximal mixing of the second and third
families of leptons: $U_{\mu 3} \cong 1/\sqrt{2}
\cong 0.7$. It is true that even in the early 1980's some grand unified
models predicted large neutrino mixing angles. (Especially
noteworthy is the remarkably prophetic 1982 paper of Harvey, Ramond, and 
Reiss,$^8$
which explicitly
predicted and emphasized that there should be
large $\nu_{\mu}- \nu_{\tau}$ mixing. However, 
in those days the top mass was expected to be 
light, and Ref. 8 chose it to be 25 GeV. That gave $V_{cb}$ in that
model to be about $0.22$. The corresponding lepton mixing
was further boosted by a Clebsch of 3. With the actual value of $m_t$
that we now know, the model of Ref. 8 would predict $U_{\mu 3}$ to
be 0.12). What makes the largeness of $U_{\mu 3}$ a puzzle
in the present situation is the fact that we now know that both
$V_{cb}$ and $m_c/m_t$ are exceedingly small.

The same puzzle exists in the context of flavor symmetry. The fact
that the quark mixing angles are small suggests that there is a
family symmetry that is only weakly broken, while the large mixings
of some of the neutrinos suggests that family symmetries are badly
broken.

The chief point of interest in looking at any model of neutrino
mixing is how it explains the large mixing of $\nu_{\mu}$ and 
$\nu_{\tau}$. This will be the feature that I will use to organize
the classification of models.

\vspace{1.0cm}

\noindent
{\bf CLASSIFICATION OF THREE-NEUTRINO MODELS}

\vspace{0.5cm}

Virtually  
all published models fit somewhere in the simple classification
now to be described.
The main divisions of this classification are based on how the
large $\nu_{\mu}-\nu_{\tau}$ mixing arises. This mixing is
described by the element $U_{\mu 3}$ of the so-called MNS matrix
(analogous to the CKM matrix for the quarks).

The mixing angles of the neutrinos are the mismatch between the
eigenstates of the neutrinos and those of the charged leptons, or
in other words between the mass matrices $L$ and $M_{\nu}$.
Thus, there are two obvious ways of obtaining large $U_{\mu 3}$:
either $M_{\nu}$ has large off-diagonal elements while $L$ is
nearly diagonal, or $L$ has large off-diagonal elements and $M_{\nu}$
is nearly diagonal. Of course this distinction only makes sense
in some preferred basis. But in almost every model there is some 
preferred basis given by the underlying symmetries of that model.
This distinction gives the first major division in the classification,
between models of what I shall call class I and class II. 
(It is also possible that the
large mixing is due almost equally to large off-diagonal elements in
$L$ and $M_{\nu}$, but this possibility seems to be realized
in very few published models. I will put them into class II.)

If the large $U_{\mu 3}$ is due to $M_{\nu}$ (class I), then it
becomes important whether $M_{\nu}$ arises from a non-see-saw
mechanism or the see-saw mechanism. We therefore distinguish
these cases as class I(1) and class I(2) respectively. In the
see-saw models, $M_{\nu}$ is given by $- N^T M_R^{-1} N$, so
a further subdivision is possible: between models in which the
large off-diagonal elements are in $M_R$ and those in which they are in
$N$. We call these class I(2A) and I(2B) respectively.

If $U_{\mu 3}$ is due to large off-diagonal elements in $L$,
while $M_{\nu}$ is nearly diagonal (class II), then the question
to ask is why, given that $L$ has large off-diagonal elements, there
are not also large off-diagonal elements in the Dirac mass matrices
of the other charged fermions, namely $U$ and $D$, causing
large CKM mixing of the quarks. In the literature there seem
to be two ways of answering this question. One way involves
the CKM angles being small due to a cancellation between large angles
that are nearly equal in the up and down quark sectors. We call
this class II(1). The main
examples of this idea are the so-called ``flavor democracy models".
The other idea is that the matrices $L$ and $D^T$ (related by
unified or flavor symmetry) are ``lopsided" in such a way that
the large off-diagonal elements only affect the mixing of fermions
of one handedness: left-handed for the leptons, making $U_{\mu 3}$
large, and right-handed for the quarks, leaving $V_{cb}$ small.
We call this approach class II(2).

Schematically, one then has

\begin{equation}
\begin{array}{ll}
I & {\rm Large \; mixing \; from } \; M_{\nu}  \\
& (1) \;\; {\rm Non \; see \; saw} \\
& (2) \;\; {\rm See \; saw} \\
& \;\;\;\;\; {\rm A.  \; Large \; mixing \; from } \; M_R \\
& \;\;\;\;\; {\rm B.  \; Large \; mixing \; from } \; N \\
II & {\rm Large \; mixing \; from } \; L  \\
& (1) \;\; {\rm CKM \; small \; by \; cancellation} \\ 
& (2) \;\; {\rm lopsided } \; L.
\end{array}
\end{equation}

Now let us examine the different categories in more detail, 
giving examples from the literature.

\vspace{0.2cm}

\noindent
{\bf I(1) Large mixing from $M_{\nu}$, non-see-saw}. 

This kind of model gives a natural explanation of the discrepancy
between the largeness of $U_{\mu 3}$ and the smallness of $V_{cb}$.
$V_{cb}$ comes from Dirac mass matrices, which are all presumably
nearly diagonal like $L$, whereas $U_{\mu 3}$ comes from the matrix 
$U_{\nu}$; and since in non-see-saw models $M_{\nu}$ comes from  
models the matrix $M_{\nu}$ comes from
completely different physics than do the Dirac mass matrices
it is not at all surprising if it has a very different form
from the others, containing some large off-diagonal elements.
While this basic idea is very simple and appealing, these models
have the drawback that in non-see-saw models the
form of $M_{\nu}$, since it comes from new physics unrelated to
the origin of the other mass matrices, is highly unconstrained.
Thus, there are few definite predictions, in general, for masses
and mixings in such schemes. However, in some schemes constraints
can be put on the new physics responsible for
$M_{\nu}$. 

As we saw, there are a variety of attractive ideas for generating
a non-see-saw $M_{\nu}$ at low energy, and there are published
models of neutrino mixing corresponding to all these 
ideas.$^{9-13}$
$M_{\nu}$ comes from triplet Higgs in Ref. 9; from 
the Zee mechanism in Ref. 10; and from R-parity and lepton-number-violating
terms in a SUSY model in Ref. 11. In Ref. 12 a ``democratic form"
of $M_{\nu}$ is enforced by a family symmetry. Several other models
in class I(1) exist in the literature.$^{13}$

\vspace{0.2cm}

\noindent
{\bf I(2A) See-saw $M_{\nu}$, large mixing from $M_R$}

In these models, $M_{\nu}$ comes from the see-saw mechanism and therefore
has the form $- N^T M_R^{-1} N$. The large off-diagonal elements in
$M_{\nu}$ are assumed to come from $M_R$, while the Dirac neutrino
matrix $N$ is assumed to be nearly diagonal and hierarchical like the other
Dirac matrices $L$, $U$, and $D$. As with the models of class I(1),
these models have the virtue of explaining in a natural way
the difference between
the lepton angle $U_{\mu 3}$ and the quark angle $V_{cb}$. The
quark mixings all come from Dirac matrices, while the lepton mixings
involve the Majorana matrix $M_R$, which it is quite reasonable to
suppose might have a very different character, with large off-diagonal
elements.

However, there is a general problem with models of this type, which
not all the examples in the literature convincingly overcome.
The problem is that if $N$ has a hierarchical and nearly diagonal
form, it tends to communicate this property to $M_{\nu}$. For example,
suppose we take $N = {\rm diag}( \epsilon', \epsilon, 1) M$, with
$1 \gg \epsilon \gg \epsilon'$. And suppose that the $ij^{th}$
element of $M_R^{-1}$ is called $a_{ij}$. Then the matrix
$M_{\nu}$ will have the form

\begin{equation}
M_{\nu} \propto \left( \begin{array}{ccc}
\epsilon^{\prime2} a_{11} & \epsilon' \epsilon a_{12} &
\epsilon' a_{13} \\ \epsilon' \epsilon a_{12} & \epsilon^2 a_{22} &
\epsilon a_{23} \\ \epsilon' a_{13} & \epsilon a_{23} & 
a_{33} \end{array} \right). 
\end{equation}

\noindent
If all the non-vanishing elements $a_{ij}$ were of the same
order of magnitude, then obviously $M_{\nu}$ is approximately
diagonal and hierarchical. 
The contribution to the leptonic angles coming from $M_{\nu}$ would
therefore typically be proportional to the small parameters $\epsilon$
and $\epsilon'$. This suggests that to 
get a value of $U_{\mu 3}$ that is of order 1,
it is necessary to have the small parameter coming from $N$ get
cancelled by a correspondingly large parameter from $M_R^{-1}$.
The trouble is that to have such a conspiracy between the magnitudes
of parameters in $N$ and $M_R$ is unnatural, in general, since these
matrices have very different origins. This problem has been 
pointed out by various authors.$^{14}$ We shall call it the
Dirac-Majorana conspiracy problem.

There are several models in the literature that fall into 
class I(2A).$^{15-17}$
Of these, an especially interesting paper is that of Jezabek and
Sumino,$^{15}$ because it shows that a Dirac-Majorana conspiracy can be
avoided. Jezabek and Sumino consider the case that the
Dirac and Majorana matrices of the neutrinos have the forms

\begin{equation}
N = \left( \begin{array}{ccc} x^2 y & 0 & 0 \\ 0 & x & x \\
0 & O(x^2) & 1 \end{array} \right) m_D, \;\;\;
M_R = \left( \begin{array}{ccc} 0 & 0 & A \\ 0 & 1 & 0 \\
A & 0 & 0 \end{array} \right) m_R,
\end{equation}

\noindent
where $x$ is a small parameter. If one computes $M_{\nu}
= -  N^T M_R^{-1} N$ one finds that

\begin{equation}
M_{\nu} = - \left( \begin{array}{ccc} 0 & O(x^4 y/A) & x^2 y/A \\
O(x^4 y/A) & x^2 & x^2 \\ x^2 y/A & x^2 & x^2 \end{array} \right)
m_D^2/m_R.
\end{equation}

\noindent
Note that this gives a maximal mixing of the second and third families,
without having to assume any special relationship between the
small parameter in $N$ (namely $x$) and the parameter in $M_R$
(namely $A$). Altarelli and Feruglio$^{16}$ generalize this example,
showing that the same effect occurs if $M_R$ is taken to
have a triangular symmetric form. 

An interesting point about the form of $M_{\nu}$ in Eq. (4) is that
it gives bimaximal mixing. This is easily seen by doing a
rotation of $\pi/4$ in the 2-3 plane, bringing the matrix to
the form 

\begin{equation}
M_{\nu}' = \left( \begin{array}{ccc} 0 & z & z' \\ z & 0 & 0 \\
z' & 0 & 2 x^2 \end{array} \right).
\end{equation}

\noindent
In the 1-2 block this matrix has a Dirac form, giving nearly maximal
mixing of $\nu_e$.

Other published models that fall into class I(2) are given in Ref. 17.

\vspace{0.2cm}

\noindent
{\bf I(2B) See-saw $M_{\nu}$, large mixing from $N$}

At least at first glance, this seems to be a less natural approach.
the point is that
if the large $U_{\mu 3}$ is due to large off-diagonal elements in $N$,
it might be expected that the other Dirac mass matrices, $U$, $D$, and $L$,
would also have large off-diagonal elements, giving large CKM angles.
There are ways around this objection, and a few interesting models 
that fall into this class have been constructed. However, 
experience seems to show that this approach is harder to make work
than the others, and fewer models of this type exist in the 
literature.$^{18}$

\vspace{0.2cm}

\noindent
{\bf II(1) Large mixing from $L$, CKM small by cancellation}
 
If the large value of $U_{\mu 3}$ comes from large off-diagonal
elements in the mass matrix $L$ of the {\it charged} leptons,
then it is most natural to assume that the other Dirac mass
matrices have large off-diagonal elements also. Why, then,
are the CKM angles small? One possibility is that the CKM
angles are small because of an almost exact cancellation between
large angles needed to diagonalize $U$ and $D$. That, in turn,
would imply that $U$ and $D$, even though highly
non-diagonal, have nearly identical forms. This is the idea realized in
so-called ``flavor democracy" models. 

In flavor democracy models, a permutation symmetry $S_3 \times S_3$
among the left- and right-handed fermions causes the Dirac mass
matrices $L$, $D$, and $U$ to have the form

\begin{equation}
L, D, U \propto \left( \begin{array}{ccc} 1 & 1 & 1 \\
1 & 1 & 1 \\ 1 & 1 & 1 \end{array} \right).
\end{equation}

\noindent
Smaller contributions that break the permutation symmetry cause
deviations from this form. These flavor-democratic forms are of
rank 1, explaining why one family is much heavier than the others.
On the other hand, the mass matrix
of the neutrinos $M_{\nu}$ is assmed to have, by an $S_3$
symmetry acting on the left-handed neutrinos, the approximate
form

\begin{equation}
M_{\nu} \propto \left( \begin{array}{ccc} 1 & 0 & 0 \\ 0 & 1 & 0 \\
0 & 0 & 1 \end{array} \right).
\end{equation}

If $M_{\nu}$ were {\it exactly} proportional to the identity, then
the basis of neutrino mass eigenstates would be undefined, and so then
would be the MNS angles. However, once the small $S_3$-violating
effects are taken into account, a neutrino basis is picked out.
It is not surprising that, typically, the neutrino angles that are
predicted are of order unity. On the other hand, the fact that
$U$ and $D$ are nearly the same in form leads to a cancellation that
tends to make the quark mixing angles small.

Exactly what angles are predicted for the neutrinos depends on the
form of the small contributions to the mass matrices
that break the permutation symmetries. There are many 
simple forms that might be assumed, and the possibilities are
rich. There exists a large and growing literature on these models.$^{19}$

The idea of flavor democracy is an elegant one, especially in
that it uses one basic idea to explain the largeness of
the leptonic angles, the smallness of the quark angles, and the
fact that one family is much heavier than the others. On the
other hand, it requires the very specific forms given in Eqs. (6) and (7),
which come from very specific symmetries. It is in this sense 
a narrower approach to the problem of fermion masses than 
some of the others I have mentioned. 

It would be interesting to know whether models of class II(1),
in which the CKM angles are small by cancellations of large
angles, can be constructed using ideas other than flavor
democracy.

\vspace{0.2cm}

\noindent
{\bf II(2) Large mixing from ``lopsided" $L$}

We now come to what I regard as the most elegant way to explain the
largeness of $U_{\mu 3}$: ``lopsided" $L$.
The basic idea is that the charged-lepton and down-quark
mass matrices have the approximate forms

\begin{equation}
L \sim \left( \begin{array}{ccc} 0 & 0 & 0 \\ 0 & 0 & \epsilon \\
0 & \sigma & 1 \end{array} \right) m_D, \;\;
D \sim \left( \begin{array}{ccc} 0 & 0 & 0 \\ 0 & 0 & \sigma \\
0 & \epsilon & 1 \end{array} \right) m_D. 
\end{equation}

\noindent
The ``$\sim$" sign is used because in realistic models these 
$\sigma$ and $\epsilon$ entries could have additional factors
of order unity, such as from Clebsches. The fact that $L$
is related closely in form to the {\it transpose} of $D$ is
a very natural feature from the point of view of $SU(5)$
or related symmetries, and
is a crucial ingredient in this approach. 
The assumption is that $\epsilon \ll 1$, while $\sigma \sim 1$.
In the case of the charged leptons $\epsilon$ controls the
mixing of the second and third families of {\it right}-handed
fermions (which is not observable at low energies), while
$\sigma$ controls the mixing of the second and third families
of {\it left}-handed fermions, which contributes to $U_{\mu 3}$
and makes it large. For the quarks the reverse is the case
because of the ``$SU(5)$" feature: the small $O(\epsilon)$
mixing is in the left-handed sector, accounting for the smallness
of $V_{cb}$, while the large $O(\sigma)$ mixing is in the right-handed
sector, where it cannot be observed and does no harm.

In this approach the three crucial elements are these:
(a) Large mixing of neutrinos (in particular of $\nu_{\mu}$
and $\nu_{\tau}$) caused by large off-diagonal elements in
the {\it charged}-lepton mass matrix $L$; (b) this off-diagonal
element appearing in a highly asymmetric or lopsided way; and
(c) $L$ being similar to the transpose of $D$ by $SU(5)$
or a related symmetry.

To my knowledge the first place that all the elements of this
approach appear is in a paper by Babu and Barr$^{20}$
and a sequel by Barr.$^{21}$
In those papers the emphasis was
on a particular mechanism (in $SU(5)$ and $SO(10)$)
by which the lopsidedness of
$L$ and $D$ can arise.
So perhaps it was not noticed by some readers
that the scheme described 
in those papers was an instance of a more general
mechanism. 

The next time that this general idea can be found
is in three papers that appeared almost simultaneously:
Sato and Yanagida,$^{22}$ Albright, Babu, and Barr,$^{23}$
and Irges, Lavignac, and Ramond.$^{24}$

It is interesting that the same mechanism was arrived at
independently by these three groups from completely different
points of view. In Sato and Yanagida the model is based on
$E_7$, and the structure of the matrices is determined by the 
Froggatt-Nielson mechanism. In Albright, Babu, and Barr, the 
model was based on $SO(10)$, and does not use the
Froggett-Nielson approach. Rather, the constraints on the form
of the mass matrices come from assuming a ``minimal" set of Higgs
for $SO(10)$ and choosing the smallest and simplest set of
Yukawa operators that can give realistic matrices. Though both
papers assume a unified symmetry larger than $SU(5)$, in both
it is the $SU(5)$ subgroup that plays the critical role in relating
$L$ to $D^T$. The model of
Irges, Lavignac, and Ramond, like that of Sato and Yanagida, uses the
Froggatt-Nielson idea, but is not based on a grand unified
group. Rather, the fact that $L$ is related to $D^T$ follows
ultimately from the requirement of anomaly cancellation for the
various $U(1)$ flavor symmetries of the model. However, it is well known
that anomaly cancellation typically enforces charge assignments that
can be embedded in unified groups. So that even though the model does
not contain an explicit $SU(5)$, it could be said to be ``$SU(5)$-like".

In the last two years, the same mechanism has been employed
by a large number of authors using a variety of approaches.$^{25}$

\vspace{1cm}

\noindent
{\bf A PREDICTIVE SO(10) MODEL WITH LOPSIDED L}

\vspace{0.5cm}

The model that I shall now describe briefly was not constructed to
explain neutrino phenomenology; rather it emerged from
the attempt to find a realistic model of the masses of the
charged leptons and quarks in the context of $SO(10)$, In particular,
the idea was to take the Higgs sector of $SO(10)$ to be as
minimal as possible, and then to find what this implied for
the mass matrices of the quarks and leptons. In fact, in the first
paper we wrote, we did not pay any attention to the neutrino
spectrum. Then we noticed that the model in that paper actually
predicted a large mixing of $\nu_{\mu}$ with $\nu_{\tau}$ and
published a follow-up paper.$^{23}$ The reason for the large
mixing of the mu and tau neutrinos was precisely the fact that
the charged lepton mass matrix has a lopsided form.

The reason this lopsided form was built into this model (which
I shall refer to as the ABB model henceforth) was that it
was necessary to account for certain well-known features of the
mass spectrum of the quarks. In particular, the mass matrix entry
that is denoted $\sigma$ in Eq. (8) above plays three crucial
roles in the ABB model that have nothing to do with neutrino
mixing. (1) It is required to get the Georgi-Jarlskog$^{26}$ factor
of 3 between $m_{\mu}$ and $m_s$. (2) It explains the value of
$V_{cb}$. (3) It explains why $m_c/m_t \ll m_s/m_b$. Remarkably,
it turns out not only to perform these three tasks, but also gives
mixing of order 1 between $\nu_{\mu}$ and $\nu_{\tau}$. Not often
are four birds killed with one stone! 

In constructing the model, several considerations guided us.
First, we assumed the ``minimal" set of Higgs for $SO(10)$. It
has been shown$^{27}$ that the smallest set of Higgs that will allow
a realistic breaking of $SO(10)$ down to $SU(3) \times SU(2) \times U(1)$,
with natural doublet-triplet splitting,$^{28}$ consists of a single
adjoint (${\bf 45}$), two pairs of spinors (${\bf 16} + \overline{{\bf 16}}$),
a pair of vectors (${\bf 10}$), and some singlets. The adjoint, in
order to give the doublet-triplet splitting, must have a VEV proportional
to the $SO(10)$ generator $B-L$. This fact is an important constraint.
Second, we assumed that the qualitative features of the quark and lepton
spectrum should not arise by artificial cancellations or numerical
accidents. Third, we required that the Georgi-Jarlskog factor arise
in a simple and natural way. Fourth, we assumed that the entries
in the mass matrices should come from operators of low-dimension that
arise in simple ways from integrating out small representations
of fermions.

Having imposed these conditions of economy and naturalness on the model
we were led to a structure coming from just six effective Yukawa terms
(just five if $m_u$ is allowed to vanish). These gave the following
mass matrices:

\begin{equation}
\begin{array}{ll}
U^0 = \left( \begin{array}{ccc} \eta & 0 & 0 \\ 0 & 0 & \frac{1}{3}
\epsilon \\ 0 & - \frac{1}{3} \epsilon & 1 \end{array} \right) m_U, \;\;\;
& D^0 = \left( \begin{array}{ccc} 0 & \delta & \delta' \\ \delta &
0 & \sigma + \frac{1}{3} \epsilon \\ \delta' & - \frac{1}{3} \epsilon 
& 1 \end{array} \right) m_D \\
& \\
N^0 = \left( \begin{array}{ccc} \eta & 0 & 0 \\ 0 & 0 & - \epsilon \\
0 & \epsilon & 1 \end{array} \right) m_U, \;\;\; & L^0 = \left( 
\begin{array}{ccc} 0 & \delta & \delta' \\ \delta & 0 & - \epsilon \\
\delta' & \sigma + \epsilon & 1 \end{array} \right) m_D.
\end{array}
\end{equation}

\noindent
(The first papers$^{23}$ gave only the structures of the second and third
families, while this was extended to the first family in a subsequent 
paper.$^{29}$)
Here $\sigma \cong 1.8$, $\epsilon \cong 0.14$, $\delta \cong |\delta'|
\cong 0.008$, $\eta \cong 0.6 \times 10^{-5}$.
The patterns that are evident in these matrices are due to the
$SO(10)$ group-theoretical characteristics of the various Yukawa terms.
Notice several facts about the crucial parameter $\sigma$ that is
responsible for the lopsidedness of $L$ and $D$. First, if $\sigma$
were not present, then instead of the Georgi-Jarlskog factor of 3,
the ratio $m_{\mu}/m_s$ would be given by 9. (That is, the Clebsch
of $\frac{1}{3}$ that appears in $D$ due to the generator $B-L$
gets squared in computing $m_s$.) Since the large entry $\sigma$ overpowers
the small entries of order $\epsilon$, the correct Georgi-Jarlskog
factor emerges. Second, if $\sigma$ were not present, $U$ and
$D$ would be proportional, as far as the two heavier families are
concerned, and $V_{cb}$ would vanish. Third, by having $\sigma \sim
1$ one ends up with $V_{cb}$ and $m_s/m_b$ being of the same order 
($\epsilon$) as is indeed observed. And since $\sigma$ does not appear
in $U$ (for group-theoretical reasons) the ratio $m_c/m_t$ comes out
much smaller, of order $\epsilon^2$, also as observed. In fact,
with this structure, the mass of charm is predicted correctly to within
the level of the uncertainties. 

Thus, for several reasons that have nothing to do with neutrinos one
is led naturally to the very lopsided form that we found gives an
elegant explanation of the mixing seen in atmospheric neutrino data!

From the very small number of Yukawa terms, and from the fact that
$SO(10)$ symmetry gives the normalizations of these terms, and not merely
order of magnitude estimates for them, it is not surprising
that many precise predictions result. In fact there are
altogether nine predictions.$^{29}$ Some of these are post-dictions
(including the highly non-trivial one for $m_c$). But several
predictions will allow the model to be tested in the future,
including predictions for $V_{ub}$, and the mixing angles $U_{e2}$
$U_{e3}$.

In the first papers it appeared that the model only gave the small-angle
MSW solution to the solar neutrino problem. In fact, if $\eta =0$,
or if forms for $M_R$ are chosen that do not involve much mixing
of the first-family right-handed neutrino with the others, then
a very precise prediction for $U_{e2}$ results that is beautifully
consistent with the small-angle MSW solution.$^{29}$ However, in a
subsequent paper$^{30}$ we showed that for other simple forms of $M_R$
the model gives bi-maximal mixing. (This happens in a way  
similar to what we saw above in Eqs. (4) and (5) for the 
Jezabek-Sumino model.) 

For more details of the ABB model and its predictions I refer you
the papers I have mentioned.

(The classification given in this talk has been somewhat expanded
in a paper by Barr and Dorsner.$^{31}$ That paper also contains a 
much more complete listing of three-neutrino models that have been 
published in the last few years. It also gives a general discussion
of expectations for the parameter $U_{e3}$.)

\vspace{1cm}

\noindent
{\bf REFERENCES}

\begin{enumerate}

\item J.W.F. Valle, Neutrino physics at the turn of the millenium
(hep-ph/9911224); S.M. Bilenky, Neutrino masses, mixings, and 
oscillations, Lectures at the 1999 European School of High Energy
Physics, Casta Papiernicka, Slovakia, Aug. 22-Sept. 4, 1999
(hep-ph/0001311).
\item M.C. Gozalez-Garcia, P.C. de Holanda, C. Pe\~{n}a-Garay, and
J.C.W. Valle, Status of the MSW solutions to the solar
neutrino problem, hep-ph/9906469
\item V. Barger and K. Whisnant, Seasonal and energy dependence
of solar neutrino vacuum oscillations, hep-ph/9903262
\item M.C. Gonzalez-Garcia, talk at International Workshop on Particles
in Astrophysics and Cosmology: From Theory to Observation, Valencia,
Spain, May 3-8, 1999.
\item M. Gell-Mann, P. Ramond, and Slansky, in {\it Supergravity,
Proc. Supergravity Workshop at Stony Brook}, ed. P. Van Nieuwenhuizen
and D.Z. Freedman (North-Holland, Amsterdam, 1979); T. Yanagida,
{\it Proc. Workshop on Unified theory and the baryon number of the
universe}, ed. O. Sawada and A. Sugamota (KEK, 1979).
\item A. Zee, Phys. Lett. {\bf B93} (1980) 389; Phys. Lett. {\bf B161}
(1985) 141.
\item C. Froggatt and H.B. Nielson, Nucl. Phys. {\bf B147} (1979) 277.
\item J.A. Harvey, D.B. Reiss, and P. Ramond, Mass relations
and neutrino oscilations in an $SO(10)$ model, Nucl. Phys.
{\bf B199} (1982) 223-268.
\item R.N. Mohapatra and S. Nussinov, Bimaximal neutrino mixing
and neutrino mass matrix, Phys. Rev. {\bf D60} (1999) 013002
(hep-ph/9809415).
\item C. Jarlskog, M. Matsuda, S. Skadhauge, and M. Tanimoto,
Zee mass matrix and bimaximal neutrino mixing, Phys. Lett.
{\bf B449} (1999) 240-252 (hep-ph/9812282).
\item M. Drees, S. Pakvasa, X. Tata, T. terVeldhuis, 
A supersymmetric resolution of solar and atmospheric neutrino
puzzles, Phys. Rev. {\bf D57} (1998) 5335-5339 (hep-ph/9712392).
\item K. Fukuura, T. Miura, E. Takasugi, and M. Yoshimura, 
Maximal CP violation, large mixings of neutrinos and democratic
type neutrino mass matrix, Osaka Univ. preprint, OU-HET-326
(hep-ph/9909415).
\item G.K. Leontaris and J. Rizos, New fermion mass textures from 
anomalous $U(1)$ symmetries with baryon and lepton number conservation,
CERN-TH-99-268 (hep-ph/9909206).
\item M. Jezabek and Y. Sumino, Neutrino mixing and seesaw
mechanism, Phys. Lett. {\bf B440} (1998) 327-331 (hep-ph/9807310); 
G. Altarelli and F. Feruglio, Neutrino mass textures
from oscillations with maximal mixing, Phys. Lett. {\bf B439} (1998) 112-118
(hep-ph/9807353).
\item M. Jezabek and Y. Sumino, Neutrino mixing and seesaw
mechanism, Phys. Lett. {\bf B440} (1998) 327-331 (hep-ph/9807310).
\item G. Altarelli and F. Feruglio, Phys. Lett. {\bf B439} (1998) 112-118
(hep-ph/9807353).
\item B. Stech, Are the neutrino masses and mixings closely
related to the masses and mixings of quarks?, talk at 23rd Johns Hopkins
Workshop on Current Problems in Particle Theory: Neutrinos in
the Next Millenium, Baltimore, MD, 10-12 June 1999 (hep-ph/9909268). 
M. Bando, T. Kugo, and K. Yoshioki, Neutrino mass textures with
large mixing, Phys. Rev. Lett. {\bf 80} (1998) 3004-3007 (hep-ph/9710417).
M. Abud, F. Buccella, D. Falcone, G. Ricciardi, and
F. Tramontano, Neutrino masses and mixings in $SO(10)$,
DSF-T-99-36 (hep-ph/9911238).
\item Q. Shafi and Z. Tavartkiladze, Proton decay, neutrino
oscillations and other consequences from supersymmetric $SU(6)$ with
pseudogoldstone Higgs, BA-99-39 (hep-ph/9905202);
D.P. Roy, talk at 6th Topical Seminar on Neutrino and AstroParticle Physics,
San Miniato, Italy, 17-21 May 1999 (hep-ph/9908262).
\item M. Fukugita, M. Tanimoto, and T. Yanagida, Atmospheric
neutrino oscillation and a phenomenological lepton mass matrix,
Phys. Rev. {\bf D57} (1998) 4429-4432 (hep-ph/9709388);
M. Tanimoto, Vacuum neutrino oscillations of solar neutrinos
and lepton mass matrix, Phys. Rev. {\bf D59} (1999) 017304
(hep-ph/9807283);
H. Fritzsch and Z.-z. Xing, Large leptonic flavor mixing and
the mass spectrum of leptons, Phys. Lett. {\bf B440} (1998) 313-318
(hep-ph/9808272);
S.K. Kang and C.S. Kim, Bimaximal lepton flavor mixing and 
neutrino oscillation, Phys. Rev. {\bf D59} (1999) 091302
(hep-ph/9811379).
\item K.S. Babu and S.M. Barr, 
Phys. Lett. {\bf B381} (1996) 202 (hep-ph/9511446).
\item S.M. Barr, 
Phys. Rev. {\bf D55} (1997) 1659 (hep-ph/9607419). 
\item J. Sato and T. Yanagida, Large lepton mixing in a coset space
family unification on $E(7)/SU(5)\times U(1)^3$, Phys. Lett. {\bf B430}
(1998) 127-131 (hep-ph/9710516).
\item C.H. Albright, K.S. Babu, and S.M. Barr, 
Phys. Rev. Lett. {\bf 81} (1998) 1167 (hep-ph/9802314);
C.H. Albright and S.M. Barr, Fermion masses in $SO(10)$ with a
single adjoint Higgs field, Phys. Rev. {\bf D58} (1998) 013002
(hep-ph/9712488).
\item N. Irges, S. Lavignac, and P. Ramond, Predictions from an
anomalous $U(1)$ model of Yukawa hierarchies, Phys. Rev. {\bf D58}
(1998) 035003 (hep-ph/9802334).
\item Y. Nomura and T. Yanagida, Bimaximal neutrino mixing in $SO(10)$
(GUT), Phys. Rev. {\bf D59} (1999) 017303 (hep-ph/9807325);
Z. Berezhiani and A. Rossi, Grand unified textures for neutrino
and quark mixings, JHEP 9903:002 (1999) (hep-ph/9811447);
K. Hagiwara and N. Okamura, Quark and lepton flavor mixings
in the $SU(5)$ grand unification theory, Nucl. Phys. {\bf B548}
(1999) 60-86 (hep-ph/9811495);
G. Altarelli and F. Feruglio, A simple grand unification view
of neutrino mixing and fermion mass matrices, Phys. Lett.
{\bf B451} (1999) 388-396 (hep-ph/9812475);
K.S. Babu, J. Pati, and F. Wilczek, Fermion masses, neutrino 
oscillations and proton decay in the light of SuperKamiokande,
(hep-ph/9812538);
R. Barbieri, L.J. Hall, G.L. Kane, and G.G. Ross, 
Nearly degenarate neutrinos and broken flavor symmetry,
OUTP-9901-P (hep-ph/9901228);
K.I. Izawa, K. Kurosawa, Y. Nomura, and T. Yanagida, 
Grand unification scale generation through anomalous $U(1)$
breaking, Phys. Rev. {\bf D60} (1999) 115016 (hep-ph/9904303);
E. Ma, Permutation symmetry for neutrino and charged lepton 
mass matrices, Phys. Rev. {\bf D61} (2000) 033012 (hep-ph/9909249);
Q. Shafi and Z. Tavartkiladze, Bimaximal neutrino mixings
and proton decay in $SO(10)$ with anomalous flavor $U(1)$,
BA-99-63 (hep-ph/9910314);
P. Frampton and A. Rasin, Non-abelian discrete symmetries,
fermion mass textures and large neutrino mixing, IFP-777-UNC 
(hep-ph/9910522).
\item H. Georgi and S.L. Glashow, Phys. Lett. {\bf B86} (1979) 297.
\item S.M. Barr and S. Raby, Minimal $SO(10)$ unification,
Phys. Rev. Lett. {\bf 79} (1998) 4748-4751.
\item S. Dimopoulos and F. Wilczek, report No. NSF-ITP-82-07 (1981),
in {\it The unity of fundamental interactions} Proceedings of the 19th
Course of the International School of Subnuclear Physics, Erice, Italy, 
1981 ed. A. Zichichi (Plenum Press, New York, 1983);
K.S. Babu and S.M. Barr, Phys. Rev. {\bf D48} (1993) 5354 (hep-ph/9306242);
K.S. Babu and S.M. Barr, Phys. Rev. {\bf D50} (1994) 3529 (hep-ph/9402291).
\item C.H. Albright and S.M. Barr, 
Phys. Lett. {\bf B452} (1999) 287 (hep-ph/9901318).
\item C.H. Albright and S.M. Barr, 
minimal Higgs model, Phys. Lett. {\bf B461} (1999) 218 (hep-ph/9906297).
\item S.M. Barr and Ilja Dorsner, hep-ph/0003058.

\end{enumerate}

\end{document}